\documentclass[useAMS,usenatbib,fleqn]{mnras}
\usepackage{listings}
\usepackage{amsmath}
\usepackage{natbib}
\usepackage{graphicx}
\usepackage{color}
\usepackage{rotating}
\usepackage{nicefrac}
\usepackage{txfonts}
\usepackage{trivfloat}
\usepackage{subcaption}
\usepackage{adjustbox}
\trivfloat{listfloat}
\usepackage{longtable}
\usepackage[percent]{overpic}
\usepackage{xspace}
\usepackage{url}
\usepackage{mathrsfs}
\usepackage{amssymb}
\usepackage{mathtools}
\usepackage{enumitem}
\usepackage[percent]{overpic}
\usepackage{soul}

\usepackage{epsfig}
\usepackage{amsmath}
\usepackage{rotating}
\usepackage{natbib}
\usepackage{multirow}
\usepackage[noend]{algpseudocode}

\newcommand{\reb}{{\sc \tt REBOUND}\xspace}
\newcommand{\whfast}{{\sc \tt WHFast}\xspace}

\lstdefinestyle{customc}{
  belowcaptionskip=1\baselineskip,
  breaklines=true,
  language=C,
  showstringspaces=false,
  basicstyle=\footnotesize\ttfamily,
}

\usepackage{array}
\usepackage{appendix}
\usepackage{comment}

\include{mnras_jdf}
\def\gsim{\;\rlap{\lower 2.5pt
 \hbox{$\sim$}}\raise 1.5pt\hbox{$>$}\;}
\def\lsim{\;\rlap{\lower 2.5pt
   \hbox{$\sim$}}\raise 1.5pt\hbox{$<$}\;}

\author[Hussain \& Tamayo]{Naireen Hussain $^1$,
    Daniel Tamayo$^{2}$\thanks{NHFP Sagan Fellow}\\
$^1$ Department of Astronomy and Astrophysics, University of Toronto, Toronto, Ontario, M5S 3H4, Canada\\
$^2$ {Department of Astrophysical Sciences, Princeton University, Princeton, New Jersey 08544, United States}\\
}

\title{Fundamental Limits From Chaos On Instability Time Predictions In Compact Planetary Systems}
\date{Draft version: Dec. 17 2019}

\vspace{0.5\baselineskip}

\begin{document}
\maketitle

\begin{abstract}
Instabilities in compact planetary systems are generically driven by chaotic dynamics.
This implies that an instability time measured through direct N-body integration is not exact, but rather represents a single draw from a {\it distribution} of equally valid chaotic trajectories.
In order to characterize the `errors' on reported instability times from direct N-body integrations, we investigate the shape and parameters of the instability time distributions (ITDs) for ensembles of shadow trajectories that are initially perturbed from one another near machine precision.
We find that in the limit where instability times are long compared to the Lyapunov (chaotic) timescale, ITDs approach remarkably similar lognormal distributions with standard deviations $\approx 0.43 \pm 0.16$ dex, despite the instability times varying across our sample from $10^4-10^8$ orbits.
We find excellent agreement between these predictions, derived from $\approx 450$ closely packed configurations of three planets, and a much wider validation set of $\approx 10,000$ integrations, as well as on $\approx 20,000$ previously published integrations of tightly packed five-planet systems, and a seven-planet resonant chain based on TRAPPIST-1, despite their instability timescales extending beyond our analyzed timescale.
We also test the boundary of applicability of our results on dynamically excited versions of our Solar System.
These distributions define the fundamental limit imposed by chaos on the predictability of instability times in such planetary systems.
It provides a quantitative estimate of the instrinsic error on an N-body instability time imprinted by chaos, approximately a factor of 3 in either direction.
\end{abstract}


\def\resonantSigma{0.142}
\def\resonantMean{0.410}
\def\randomSigma{0.161} %
\def\randomMean{0.440} %


\section{Introduction}
\label{intro}

A central effort in exoplanet science is the characterization of orbital architectures in planetary systems \citep{Lissauer11, Fabrycky14}.
While precise masses and orbital eccentricities have been measured in several cases \citep[e.g.][]{Carter12, Pepe13, Hadden14, Jontof14, Hadden17}, these parameters typically have large uncertainties, especially for the small rocky planets we are often most interested in.

An additional constraint that can help narrow uncertainties comes from stability considerations.
Typical exoplanet systems are several Gyrs old, so it is natural to require that the inferred orbital parameters yield an orbital configuration that is long-term stable.
Of course planetary systems can (and likely do) undergo orbital instabilities \citep[e.g.,][]{Volk15, Pu15, Izidoro17}, but it would be unlikely to catch a planetary system just before such a cataclysm, for example in the next thousandth of its lifetime.
In exoplanetary systems, where lifetimes can approach one trillion orbital timescales, stability can therefore provide a long lever arm that can significantly constrain orbital parameter uncertainties \citep[e.g.][]{Steffen13, Tamayo14a, Tamayo15, Quarles17, Tamayo17, Wang18}.

Currently, determination of stability lifetimes for particular planetary configurations is performed through computationally expensive N-body integrations.
Much effort has been spent on faster prediction of instability times drawing from suites of N-body integrations, both through parametrized empirical fits \citep[e.g.][]{Chambers96, Yoshinaga99, Marzari02, Zhou07, Faber07, Smith09, Funk10, Pu15, Obertas17, Wu19}, analytically \citep{Quillen11}, and through machine learning techniques \citep{Tamayo16, Lam18}.

In this paper we instead examine the fundamental limits on prediction imposed by the chaotic dynamics of unstable planetary systems.
One might naively assume that prediction on chaotic systems is hopeless, but this worry is unfounded.
While individual steps in a drunkard's random walk might be unpredictable, the cumulative effect of many steps approaches a well-defined statistical distribution.
Similarly, individual chaotic orbital kicks are inherently unpredictable, but if reaching orbit-crossing configurations requires many such small kicks, the overall distribution of timescales to instability can nevertheless reach a well-defined and predictable form.

This implies that even in the limit of exquisitely precise initial conditions, a particular instability time from an N-body integration should not be viewed as the definitive answer, but rather as a single draw from a {\it distribution} of instability times corresponding to that initial orbital configuration.
It is therefore valuable to understand the shapes and properties of such instability time distributions (ITDs), which set the fundamental prediction limit not only for approximate methods like those cited above, but also for the direct N-body integrations that are currently the gold standard.

To our knowledge, the first mention of the shape of ITDs in unstable, closely packed systems is in the appendix of \cite{Chatterjee08}.
They found distributions with the logarithm (all logarithms in this paper are base 10) of instability times rising exponentially to a maximum, and then falling off linearly.
By contrast, \cite{Rice18} report lognormal shapes of ITDs across a small sample of compact systems.
In this paper we systematically explore a large sample of systems, and provide a physical interpretation to account for this apparent contradiction.
Our empirical characterization for the ITDs of closely packed planetary systems provides the fundamental limit against which instability time prediction methods should be compared, and provide `error bars' for instability times quoted from direct N-body integrations.

The paper is organized as follows.
In Sec.\ref{methods} we describe the dataset of N-body integrations we consider.
In Sec.\ref{results} we characterize the shapes of ITDs and interpret them in the context of chaotically diffusing dynamical systems.
We provide a numerical and analytic approximation to ITDs across compact multi-planet systems with a broad range of parameters, and test those models against a large sample of N-body integrations.
In Sec.\:\ref{generalization} we apply our ITDs to integrations of different types of planetary systems to explore their range of applicability, and we conclude in Sec.\:\ref{conclusion}.

\section{Methods}
\label{methods}

In a companion paper focused on predicting instability times in closely packed planetary systems, Tamayo et al. ({\it in prep.}), henceforth T+, generated and ran large numbers of N-body integrations of three-planet systems spanning one billion orbits of the innermost planet.
We focus on the three-planet case because systems of three or more planets exhibit qualitatively similar behavior in instability times \citep{Chambers96}.
In the two-planet case, there are enough conserved quantities to provide powerful analytical constraints on stability \citep{Wisdom80, Marchal82, Gladman93, Deck15, Hadden18}.
The result is that outside a narrow band in phase space where the dynamics are chaotic but the motions are constrained to avoid close encounters, instabilities either typically happen promptly (on the synodic timescale on which the inner planet overtakes its outer neighbor) or don't happen at all.
In such short random walks, we do not expect ITDs to settle to well defined statistical distributions.
By contrast, instabilities in systems of three or more planets can span many decades in time and we will find that their ITDs do converge to uniform shapes.

The dataset of T+ was focused on typical parameters observed in compact multi-planet systems discovered by Kepler.
In particular, adjacent pairs planets are constrained to being less than 30 mutual Hill radii apart, which approximately corresponds to being more dynamically packed than our Solar System's terrestrial planets.
Planets' mass ratios with the central star span that of $\sim$ Mars-Neptune relative to the Sun, and orbits are initially eccentric and inclined.
More details are provided in Appendix \ref{dataset}.

We first drew a subset (detailed below) from this wider set of initial conditions for analysis.
For each of these starting configuration, we integrated 500-1000 shadow trajectories (realizations), where we randomly perturbed the middle planet's initial position by one part in $10^{11}$.
Because the dynamics are chaotic, each of these shadow trajectories generally yields a different, equally valid, draw from the initial condition's ITD.
For computational reasons, we ran the shadow trajectories for $10^8$ orbits, ten times shorter than the integrations of T+.

To choose the subsample for our numerical experiment, we focused on initial conditions that had nominal instability times (from T+) in the range of $10^4-10^7$ orbits.
This was because we found empirically that shadow trajectories in many systems with instability times shorter than $10^3$ orbits often did not have time to separate from the nominal ones before collision.
This resulted in a single spike of instabilities all at the same time.
We therefore chose a conservative lower boundary of $10^4$ orbit instability timescales.
On the high end, because we ran our shadow integrations out to $10^8$ orbits, we made a cut at $10^7$ orbits in order to mitigate the number of shadow trajectories that would reach our arbitrary boundary.
We check below that this was indeed the outcome.

Following T+, we also considered and compared two populations of initial conditions, one with randomly assigned orbital angles and eccentricities (henceforth. the `random' dataset), the other where one pair of planets is initialized in or near a strong mean motion resonance (MMR), and the remaining planet is randomly initialized (henceforth the `resonant' dataset).
This is an important comparison given that the dynamical behavior can be different between the two cases, and \cite{Chatterjee08} found different ITD shapes when comparing systems close to and far from strong MMRs.
We provide details of the setup in Appendix \ref{dataset}.

\section{Results}
\label{results}

\subsection{Instability Time Distributions}

\begin{figure}
    \centering
    \includegraphics[width=0.5\textwidth,  clip= true, trim={0cm 0cm 0cm 0cm }]{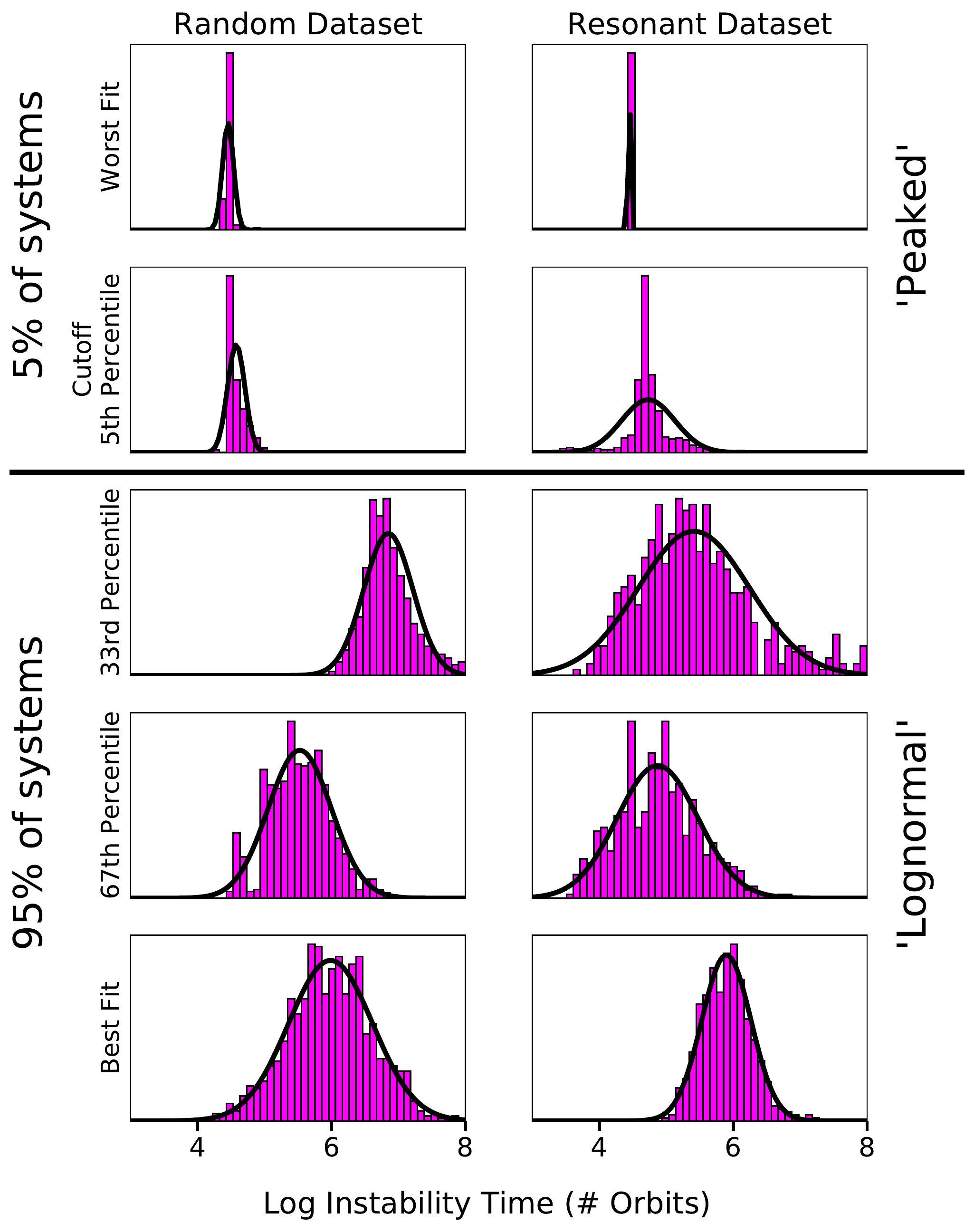}
    \caption{Instability time distributions (ITDs) for 10 selected systems in our analyzed set. Most have approximately lognormal shapes. Histograms are organized from least lognormal (top) to most lognormal (bottom), with the goodness of fit percentile labeled on the left axis. The sample of closely packed, randomly initialized systems are in the left column, systems initialized in and near strong mean motion resonances are in the right column. We find that both datasets are roughly consistent with one another statistically, and interpret the evolution from top to bottom of the plot as resulting from longer random walks to instability, which allow the ITDs to settle to lognormal shapes. We reject the worst 5\% of systems (above the black line) from our analysis (see text).}
    \label{hists}
\end{figure}

For both the random and resonant datasets, examination of the distributions of instability times for different initial conditions reveals a dichotomy.
We find that in most cases, the distribution of instability times can be approximately modeled as a lognormal distribution (e.g., the bottom rows of Fig.\:\ref{hists}), in agreement with \cite{Rice18}.
Some initial conditions, however, have `peaked' distributions of instability times with rapidly decaying tails (top row of Fig.\:\ref{hists}).

To better quantify this statement, we fit lognormal distributions using Markov Chain Monte Carlo with the {\sc \tt emcee}\xspace package \citep{Foreman13}.
We then separated the two populations by applying the Kolmogorov-Smirnov (KS) test between the observed distribution of instability times and an equal number of samples drawn from the best-fit lognormal.

We find that for both the resonant and random datasets, there is a gap in the distribution of KS-test p-values that robustly separate the datasets into two well-defined populations.
Because the random initial conditions have two times more integration samples than the resonant cases and can therefore better be distinguished from their best-fit lognormals, the threshold p-value separating `lognormal' from `peaked' distributions differs in the two cases.
For a more one-to-one comparison we therefore split the two datasets into percentiles based on their KS-test p-values.
We find that in both cases, the gap in the distribution of p-values between `lognormal' and `peaked' fits occurs approximately at the 5th percentile.
We plot various representative percentiles, including this cutoff, in Fig.\:\ref{hists}, from the worst-fit (most `peaked') distribution to the best-fit 'lognormal'.

From this exercise, we conclude that there is no strong qualitative difference for the distribution of instability times of randomly generated closely-packed systems and systems with pairs of planets initialized in and near strong MMRs.
We will corroborate this conclusion more quantitatively in Sec.\:\ref{widthsec} and in Sec.\:\ref{uncertainty} on the much wider sample of Tamayo et al.
Finally, we note that at some level the consistency between the random and resonant datasets is coincidental.
If we quantitatively understood the process that generates `peaked' distributions, we could generate a dataset only of initial conditions that would generate such distributions.
Nevertheless, given the fact that the parameters used to generate the original random and resonant distributions were only tweaked so as to generate comparable numbers of stable and unstable systems over $10^9$ orbits, the rarity of the `peaked' distributions implies they do not occupy a sizeable fraction of the phase space, and in the following section we provide a physical metric for identifying them.

\subsection{The Road To Instability}
Chaos in Hamiltonian systems generically arises from resonance overlap \citep{Chirikov79}.
Planetary orbits become orbit-crossing and destabilize due to the subsequent chaotic transport of their actions, in particular those associated with the orbital eccentricities.
In perhaps the simplest model, one can consider their diffusion under a random walk, where random steps are taken on the decoherence timescale (i.e., the Lyapunov e-folding time on which nearby trajectories diverge exponentially), and the step size in action space is given by the width of the overlapping resonances \citep{Murray97}.

We interpret the sequence of histograms in Fig.\:\ref{hists} progressing from sharply peaked to lognormal as roughly reflecting random walks of increasing length.
Different walkers (realizations) on random walks that reach instability within less than a few steps (Lyapunov times) will not have time to separate from one another and will all reach instability at roughly the same time.

By contrast, the more steps required to reach an orbit-crossing configuration, the more different walkers will separate, and the better they will approach a well defined statistical distribution of instability times.

To test this interpretation, we numerically measured the Lyapunov time from a single realization of each initial condition in our dataset.
We then used each integration to calculate the number of Lyapunov times that elapse before instability is reached for each of the systems in our sample.

This simple picture is complicated by the fact that dynamical systems, particularly unstable ones, do not have a single Lyapunov time.
As such systems approach orbit-crossing configurations, the Lyapunov time can shorten dramatically.
As a simple prescription, we therefore choose to measure the Lyapunov time over the first tenth of the timescale to instability, though we checked that the exact fraction adopted does not significantly affect the results.

\begin{figure}
    \centering
    \includegraphics[width=0.5\textwidth,  clip= true, trim={0cm 0cm 0cm 0cm }]{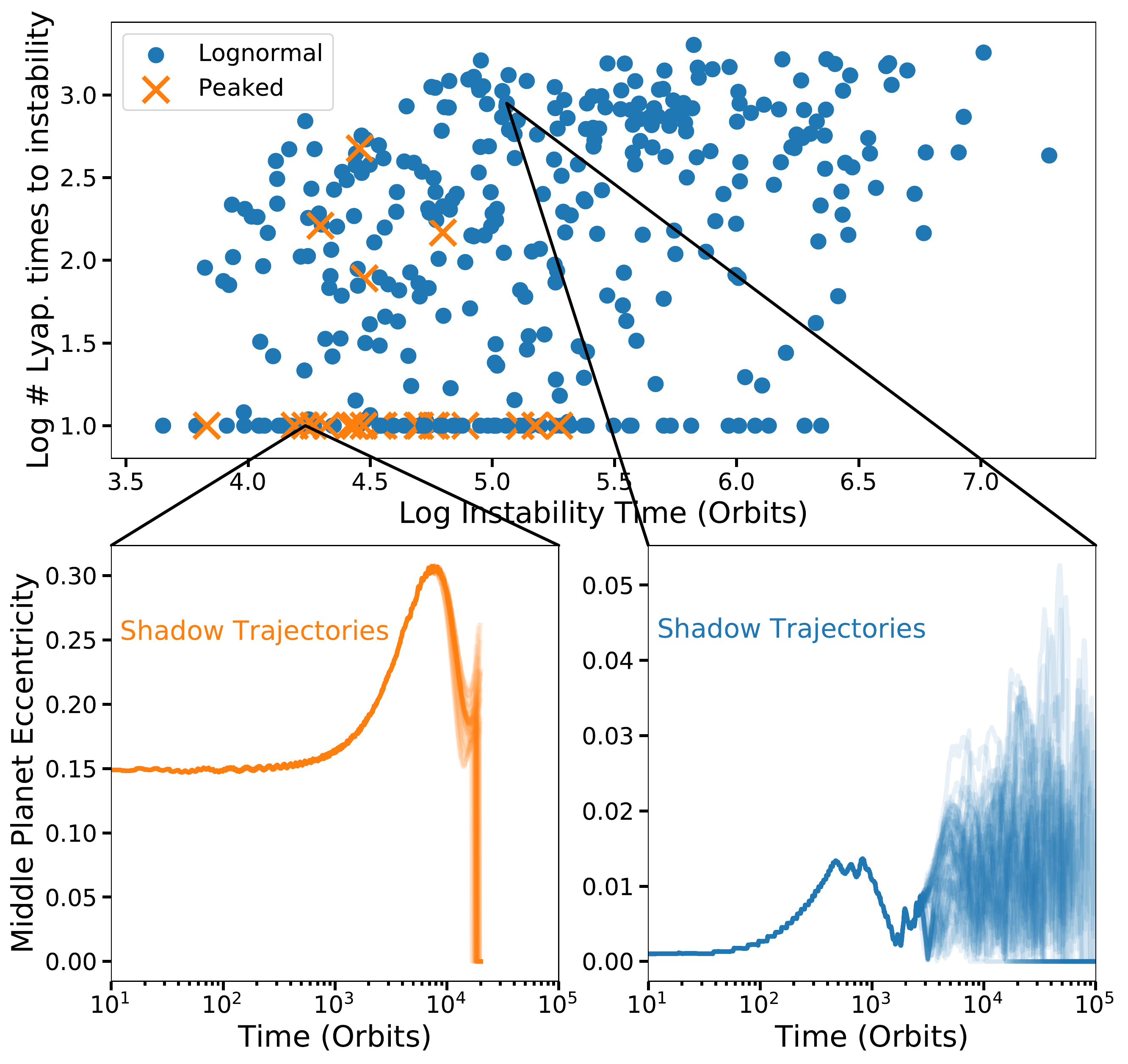}
    \caption{Number of chaotic (Lyapunov) timescales to instability vs. instability time, for the combined random and resonant datasets. `Lognormal' distributions retained for analysis (below the horizontal black line in Fig.\:\ref{hists}) are shown as blue circles, rejected `peaked' distributions are shown as orange Xs. The plot shows that most of the sharply peaked distributions (see Fig.\:\ref{hists}) correspond to systems consistent with non-chaotic dynamics. Such systems are driven to instability by simple dynamics rather than any chaotic diffusion process, so shadow trajectories all collide at essentially the same time. An example is shown in the bottom left panel, overplotting the middle planet's eccentricity evolution for all shadow trajectories. They all reach instability (where they sharply fall to zero) at nearly the same time, before they have time to diverge. Bottom right panel shows an example yielding a lognormal distribution, which shows vigorous mixing of shadow trajectories before instability.}
    \label{outliers}
\end{figure}

We plot the result in the top panel of Fig.\:\ref{outliers}, finding that the systems with sharply peaked ITDs (orange) are indeed typically ones that don't survive more than a few Lyapunov times.
These are cases where the resonant or secular dynamics {\it already} drives them onto crossing orbits, without the need for any additional chaotic diffusion.
An example such system is shown in the bottom left panel of Fig.\:\ref{outliers}.
We overplot the middle planet's eccentricity evolution for each of fifty realizations, showing that different shadow trajectories do not have time to diverge before crossing orbits (where the lines artificially drop to zero).
By contrast, systems yielding lognormal distributions in instability times show vigorous mixing of shadow trajectories (example shown in the bottom right panel of Fig.\:\ref{outliers}).

Of course, the boundary between these two behaviors is not perfectly sharp, which explains why four of the orange points in Fig\:\ref{outliers} are mixed with the blue ones.
For example, the worst outlier corresponds to the right panel in the second row of Fig.\:\ref{hists}, at the boundary we imposed between peaked and lognormal distributions.
Nevertheless, the populations are sufficiently different that the separation into two distinct classes is conceptually and qualitatively useful.
Five sixths of our peaked distributions show no measurable chaotic divergence of nearby orbits in our short integrations and fall along the bottom of Fig.\:\ref{outliers}.

This lower boundary at 10 Lyapunov times is an artificial result of our measured Lyapunov times defaulting to the length of the integration (one tenth of the instability time) when we measure no chaotic divergence of nearby trajectories over that timescale.
The fact that we need to measure the Lyapunov time over a timescale short compared to the instability time (in order to avoid an artificially short Lyapunov time from the period before instability) means that we don't resolve the Lyapunov time for a small number (also $\approx 1/6$) of the lognormal distributions, i.e., the. blue points along the bottom of Fig.\:\ref{outliers}.
In over 80\% of those cases, the nominal instability time (i.e., the time measured in the first integration) is unluckily short, in the sense that the mean of the ITD is longer and the nominal instability time drew a shorter than average value.
This makes it less likely that an integration over a tenth of this shortened instability time estimate will measure chaotic divergence of nearby orbits.
Using the mean of the ITD instead of the instability time removes several of these false positives, but we typically do not have access to the mean ITD value.

Our practical recommendation is to run a single N-body integration, and then run a second integration over a tenth the instability time to measure the Lyapunov time, which is computationally negligible.
Measuring more than ten Lyapunov times to instability is strong indication that the system goes unstable through chaotic transport, and follows the distributions given in this work.
Not measuring chaotic divergence is ambiguous and requires more computation, but running even a handful of shadow integrations should easily disambiguate between the sharply peaked and lognormal distributions illustrated in Fig.\:\ref{hists}.

We also see from Fig.\:\ref{outliers} that for instability times $\gtrsim 10^5$ inner-planet orbits ($\approx 63\%$ of our sample), systems almost invariably exhibit lognormal distributions of instability times (only $\approx 1\%$ are peaked vs. $\approx 13\%$ below $10^5$ orbits).
We suspect this boundary roughly represents the longest Lyapunov time achievable for our particular sample of closely spaced planets.
Chaos causes nearby trajectories to diverge due to repeated encounters with the unstable fixed points in phase space created by resonances, and these are encountered $\sim$ once per libration timescale \citep{Murray97}.
Lyapunov times should therefore scale with resonant libration timescales, which in particular scale inversely with planetary mass.
Indeed, the rightmost `peaked' distribution (orange X) in Fig.\:\ref{outliers} with a log instability timescale of $\sim 5.5$ is at the low-mass end of our distribution.
Planet mass ratios (with the central star) were drawn independently and log-uniformly in the range $[10^{-7}, 10^{-4}]$, and this system had mass ratios of $5\times 10^{-7}, 2 \times 10^{-6},$ and $2\times 10^{-7}$.
We expect that sharply peaked instability time distributions would extend to longer instability times (right) in Fig.\:\ref{outliers} for lower-mass systems than probed in our sample, and vice-versa.

In particular, this can explain the discrepancy between the asymmetric ITDs reported by \cite{Chatterjee08} and the lognormals found by \cite{Rice18}.
The former study considered closely packed gas giants at large separations from their host star with very short instability times peaked between $\sim 1-100$ orbits.
In this regime one takes very few steps to instability; in fact, close encounters even beyond the planets' Hill spheres can be sufficient for ejection from the system.
By contrast, the lower mass planets considered here and by \cite{Rice18} must take many steps to reach instability, and settle to well defined, predictable statistical distributions.
This is a fortunate state of affairs.
The unpredictable walks are short and trivial to integrate numerically, while the long walks that can be computationally prohibitive reach sharply peaked lognormal distributions of instability times.
We defer a detailed quantitative model to future work, and focus instead on a statistical description of these distributions of instability times to identify the features that a successful model should explain.

\subsection{A Universal Width For the Lognormal Distribution of Instability Times in Closely Packed Planetary Systems?} \label{widthsec}

Having found a practical method for identifying outlier initial conditions with sharply peaked distributions of instability times, we now focus on characterizing the limiting lognormal distributions that we argue are plausible outcomes of long random walks with many steps.

\begin{figure}
    \centering
    \includegraphics[width= 0.5\textwidth]{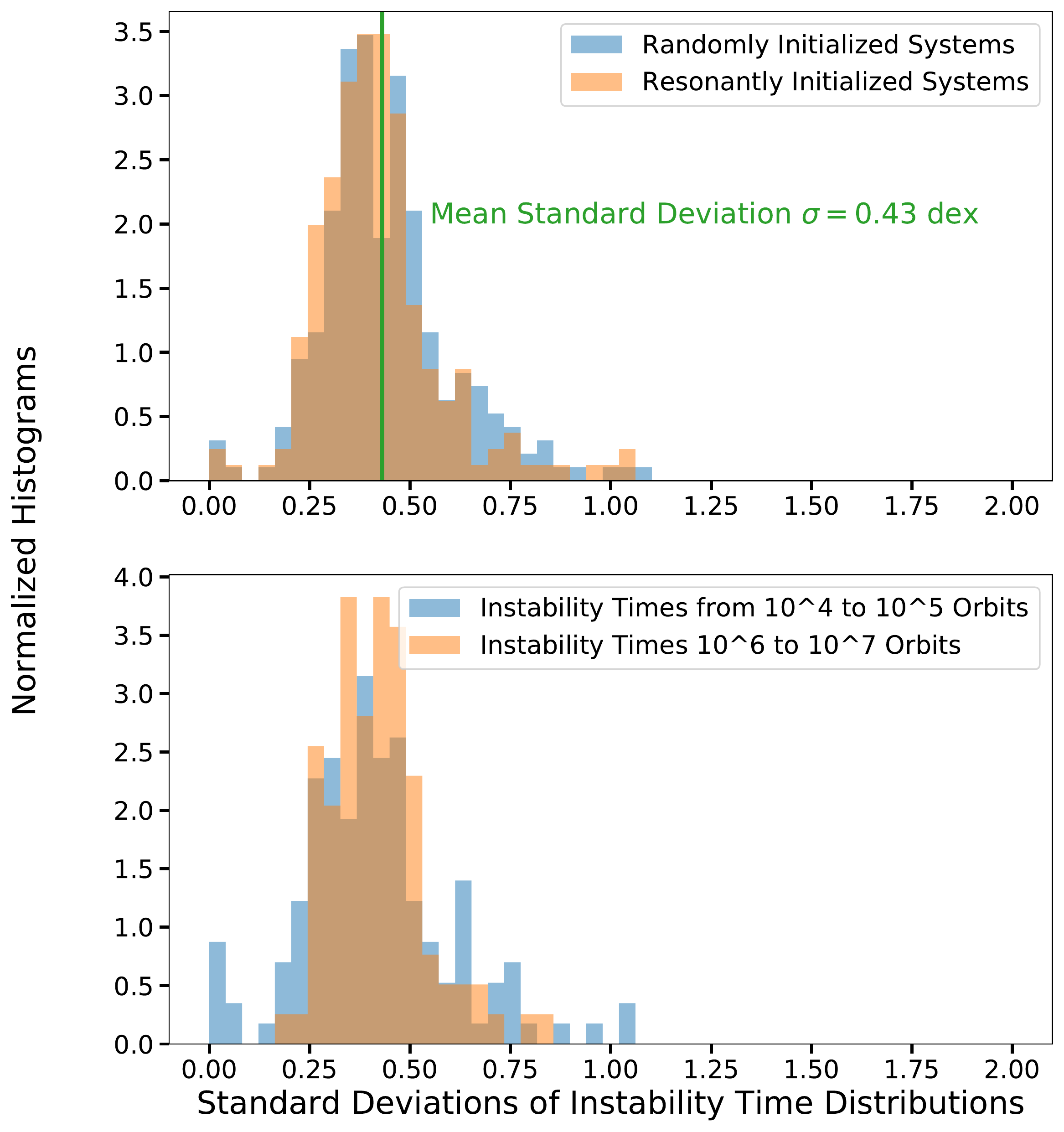}
    \caption{Top panel compares the distribution of standard deviations of the lognormal ITD fits (black lines in each panel of Fig.\:\ref{hists}) between the randomly and resonantly initialized sets of systems, showing them to be qualitatively consistent and to peak sharply at $\approx 0.43$ dex${ \pm 0.16}$. Bottom panel combines random and resonant datasets, and looks at the instability distributions at either end of the tails, showing that the standard deviations are independent of the mean instability time over the range we sampled.}
    \label{fig:comparing_sigmas}.
\end{figure}

While predicting the mean of these distributions is a difficult and unsolved problem \citep[e.g.,][]{Chambers96, Tamayo16, Obertas17}, we can see in Fig.\:\ref{hists} that the {\it standard deviations} of the lognormal distributions tend to be similar to one another.
More quantitatively, in the top panel of Fig.\:\ref{fig:comparing_sigmas}, we plot the distribution of these standard deviations from the collection of best-fit lognormals like those shown in black lines in Fig.\:\ref{hists}.
We find that the standard deviations are clustered around $\approx 0.43$ dex$ \pm 0.16$

In addition, the distribution of lognormal standard deviations in the resonant and random datasets are qualitatively similar to one another. A KS-test between the two distributions yields a p-value of 0.06, marginally consistent with having been drawn from the same distribution\footnote{The two distributions consistently return KS p-values above 0.05 even if we quadruple the percentile at which we set the threshold between `lognormal' and `peaked' systems in Fig.\:\ref{hists}.}.
This may be the result of the fact that in the tightly packed regime with period ratios $<1.3-1.5$, even our `random' systems that are not specifically initialized in resonant islands always have nearby strong first and second-order MMRs.

We also note that this contrasts with the results of \cite{Chatterjee08}, who find significantly different ITDs for systems near strong MMRs.
This is likely the result of their probing very short random walks with instability times much shorter than our lower cutoff at $10^4$ orbits; it may be that temporary phase protection from strong MMRs may skew the distributions.

Our lognormal standard deviations are also about a factor of two larger than the standard deviations found for the three initial conditions considered by \cite{Rice18}.
We suspect this is the result of their adopting effectively planar initial conditions, where the vertical orbital excursions are much smaller than the Hill radii sizes.
Indeed, \cite{Rice18} find that the systems' behavior changes sharply at this transition from an effectively 2-D to 3-D geometry \citep[see also Sec.\:3.5 of][]{Tamayo15}).
We speculate that in addition to shortening instability times \citep{Rice18}, chaotic diffusion in a lower-dimensional 2-D geometry also leads to a smaller {\it spread} in instability times.
We test this on a set of 20,000 co-planar, initially circular integrations of five planets in Sec.\:\ref{obertas}, and indeed find good agreement there with the results of \cite{Rice18}.

Perhaps most importantly, we find that the ITD standard deviations are independent of the mean instability time, which in our sample varies over $[10^4, 10^7]$ orbits.
To test this we combined the random and resonant samples, and split the combined dataset to examine the tails of the distribution. We compare systems with short  instability times (mean in the range of $[10^4, 10^{5}]$ orbits), and long instability times ($[10^6, 10^{7}]$ orbits). We plot the two distributions of standard deviations in the bottom panel of Fig.\:\ref{fig:comparing_sigmas}, and find that a KS test between the two distributions yields a p-value of $\approx 0.2$, consistent with having been drawn from the same distribution.
This is an important feature for any dynamical model for such instabilities to match.
We test the boundaries of applicability of this result in Sec.\:\ref{generalization}, but first test it on the wider dataset of Tamayo et al. ({\it in prep}).

\subsection{How uncertain are instability times from N-body integrations?} \label{uncertainty}

The results above imply that if one runs many shadow trajectories around a given tightly packed configuration, one would expect a lognormal ITD with standard deviation $\approx 0.4$ dex; however, this is not the typical situation.
Being computationally constrained, one usually runs a single N-body integration for each initial condition.
We can now statistically quantify the uncertainties in such instability time measurements.

One might first ask what one expects the mean $\mu$ of the ITD to be, given a single measurement (draw) of the instability time\footnote{For simplicity of notation, all times in this section are logarithmic so the lognormal distributions are simply Gaussians.} $t$ from an N-body integration.
In the agnostic case, with flat priors on $t$ and $\mu$, the probability of $\mu$ given $t$, $P(\mu|t)$, is simply equal to $P(t|\mu) \propto \exp\left[-(t-\mu)^2/(2\sigma^2)\right]$.
In other words, given a single instability time $t$ from an N-body integration, the best estimate on {\it the mean} $\mu$ of the ITD is $t$ itself, and it is distributed normally around $t$ with a standard deviation $\sigma \approx 0.4$.

One could also ask what instability time $t'$ one would expect if one were to run an additional N-body integration with effectively the same initial conditions (offset by machine precision).
This is a slightly different question.
In this case $t'$ would be drawn from the same Gaussian distribution as $t$ with mean $\mu$ and standard deviation $\sigma$, but the complication is that we have imperfect information on where $\mu$ is from our single measurement $t$; it is distributed normally around $t$ as discussed in the previous paragraph.
Combining many lognormals of standard deviation $\sigma$ with different possible values of $\mu$ should yield a wider distribution with larger uncertainties.

A simple way to evaluate this integral is to realize that we are after the distribution of $t'-t$, where $t'$ and (negative) $t$ are drawn independently from Gaussians with mean $\mu$ and $-\mu$, respectively, and standard deviations $\sigma$.
The resulting distribution for $t'-t$ is then also Gaussian with the sum and mean of the variances, i.e., zero mean, and variance $2\sigma^2$.
In other words, when an N-body integration measures an instability time $t$, running a second shadow integration will yield an instability time lognormally distributed around $t$, with standard deviation $\sqrt(2) \sigma$, or $\approx 0.6$ dex (since $\sigma \approx 0.4$ dex from Fig.\:\ref{fig:comparing_sigmas}).

This wider distribution reflects the uncertainty on the mean of the ITD, stemming from the single prior instability time measurement $t$.
Of course, one can always run more shadow integrations to reduce the uncertainties on $\mu$; in the limit where $\mu$ is known, it reduces to the case above where uncertainties on future draws $t'$ yields $\sigma \approx 0.4$, i.e., the standard deviation physically imprinted by the chaotic dynamics.

We test these predictions, derived from the ITDs of $454$ configurations across the random and resonant datasets analyzed above, on the wider sample of Tamayo et al. ({\it in prep.}).
Each of the integrations in the wider dataset do not have 1000 shadow trajectories like the cases we analyzed in this paper.
However, Tamayo et al.({\it in prep}) did run a {\it single} additional `shadow' integration for each of their approximately 30,000 initial conditions, providing a second draw from the ITD for comparison with the prediction.

In this wider dataset, we do not have measured Lyapunov times to filter out peaked ITDs (which have much narrower standard deviations, cf. top panels in Fig.\:\ref{hists}).
But from Fig.\:\ref{outliers}, we can see that most outliers are avoided beyond instability times of $\sim 10^5$ orbits.
We therefore select systems that went unstable between $10^5-10^8$  orbits (with the upper limit chosen so that our halting the integrations at $10^9$ does not have a strong effect), as compared to our sample analyzed above with instability times between $10^4-10^7$ orbits.
This left us with 8673 test systems from this wider sample, 4496 random systems and 4177 resonant systems.

\begin{figure}
    \centering
    \includegraphics[width= 0.5\textwidth]{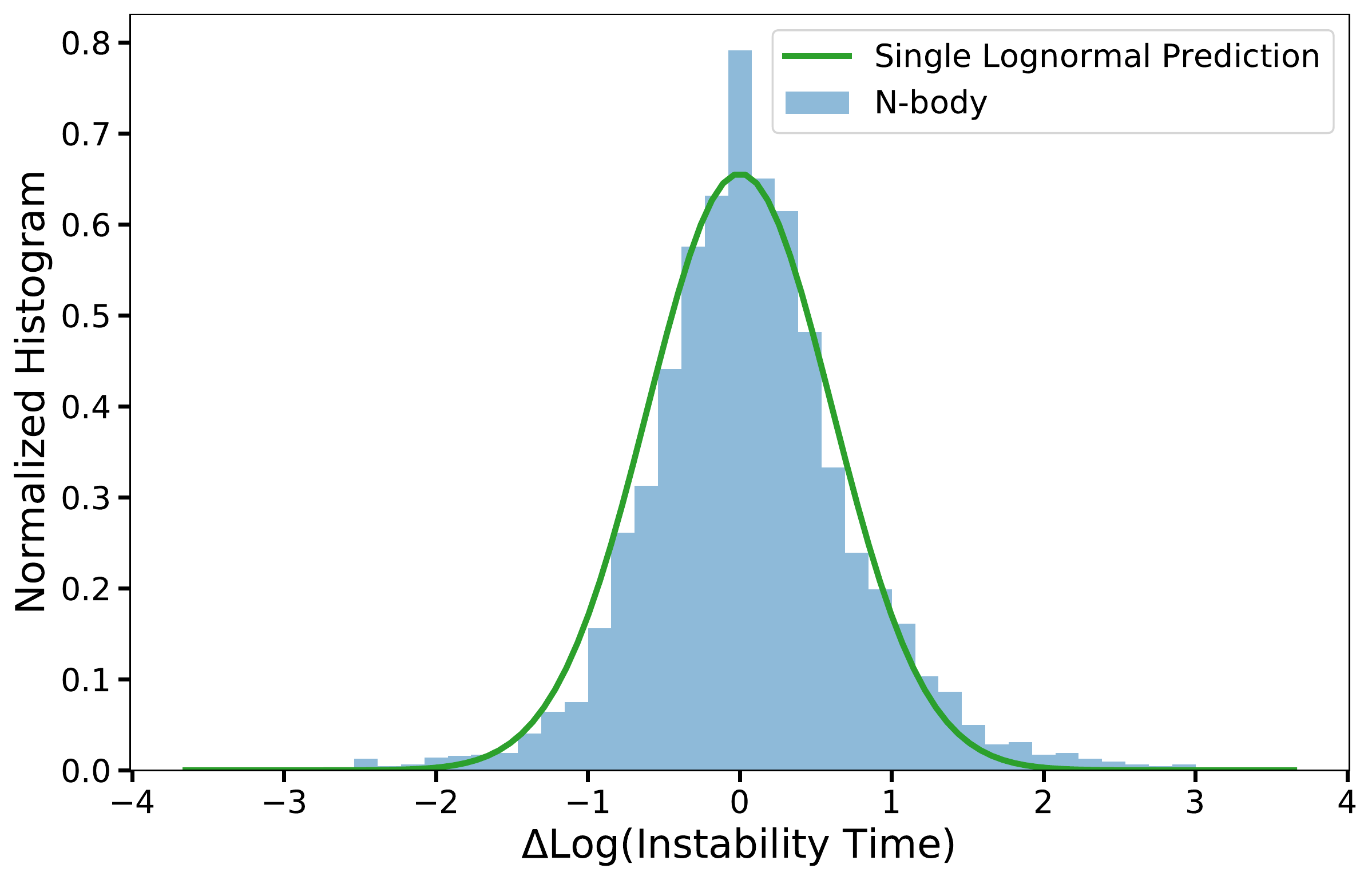}
    \caption{Histogram of the differences in log instability times measured between 8673 pairs of shadow integrations with instability times between $10^5-10^8$. The green line is the prediction from our detailed characterization of a much smaller subset of 454 systems above (see text).}
    \label{fig:comparing_shadow_systems}.
\end{figure}

We show the result in Fig.\:\ref{fig:comparing_shadow_systems}.
As argued above, we expect a Gaussian distribution with standard deviation $\sqrt{2}\sigma$.
In principle, $\sigma$ should be drawn from the distribution of standard deviations in Fig.\:\ref{fig:comparing_sigmas}).
But given the strong clustering in standard deviations, we choose to simply adopt the mean standard deviation across the 454 ITDs we analyzed of $\sigma = 0.43$ dex (green vertical line in Fig.\:\ref{fig:comparing_sigmas}).
We plot this simple lognormal prediction in green in Fig.\:\ref{fig:comparing_shadow_systems}.
Despite this larger sample including many systems ($\approx 1/4$) with instability times between $10^7-10^8$ orbits, a range poorly sampled by our analysis above, we nevertheless find excellent agreement.
We attribute the excess overpopulating the center of the distribution to peaked distributions not caught by our simple cut at $10^5$ orbits.

In summary, assuming that all lognormal ITDs have a single standard deviation (green curve) gives a good qualitative fit across the phase space of tightly packed systems of three planets.
This provides a useful and conservative rule of thumb when discussing uncertainties in instability times from N-body integrations, which we summarize in Sec.\:\ref{conclusion}.
In the next section we test our predictions on a wider variety of systems to test their limits of applicability.

\section{Generalization} \label{generalization}

While we showed in Sec.\:\ref{uncertainty} that our results from a small subset of initial configurations generalized well to a much larger sample, the initial conditions were statistically generated in the same way (see Sec.\:\ref{methods}).
We now test these results on other types of planetary systems to check their generalizability and limits of applicability.

\subsection{Equally spaced, five-Earth systems} \label{obertas}

\cite{Obertas17} ran approximately $20,000$ integrations of five co-planar and equally-spaced Earth-mass planets around solar-mass stars, on initially circular orbits.
They varied only the separation between adjacent pairs of planets, and randomly sampled the planets' initial orbital phases.
They consider like us the tightly packed regime, but the number of planets and initial configurations are significantly different to our dataset.

\begin{figure}
    \centering
    \includegraphics[width= 0.5\textwidth]{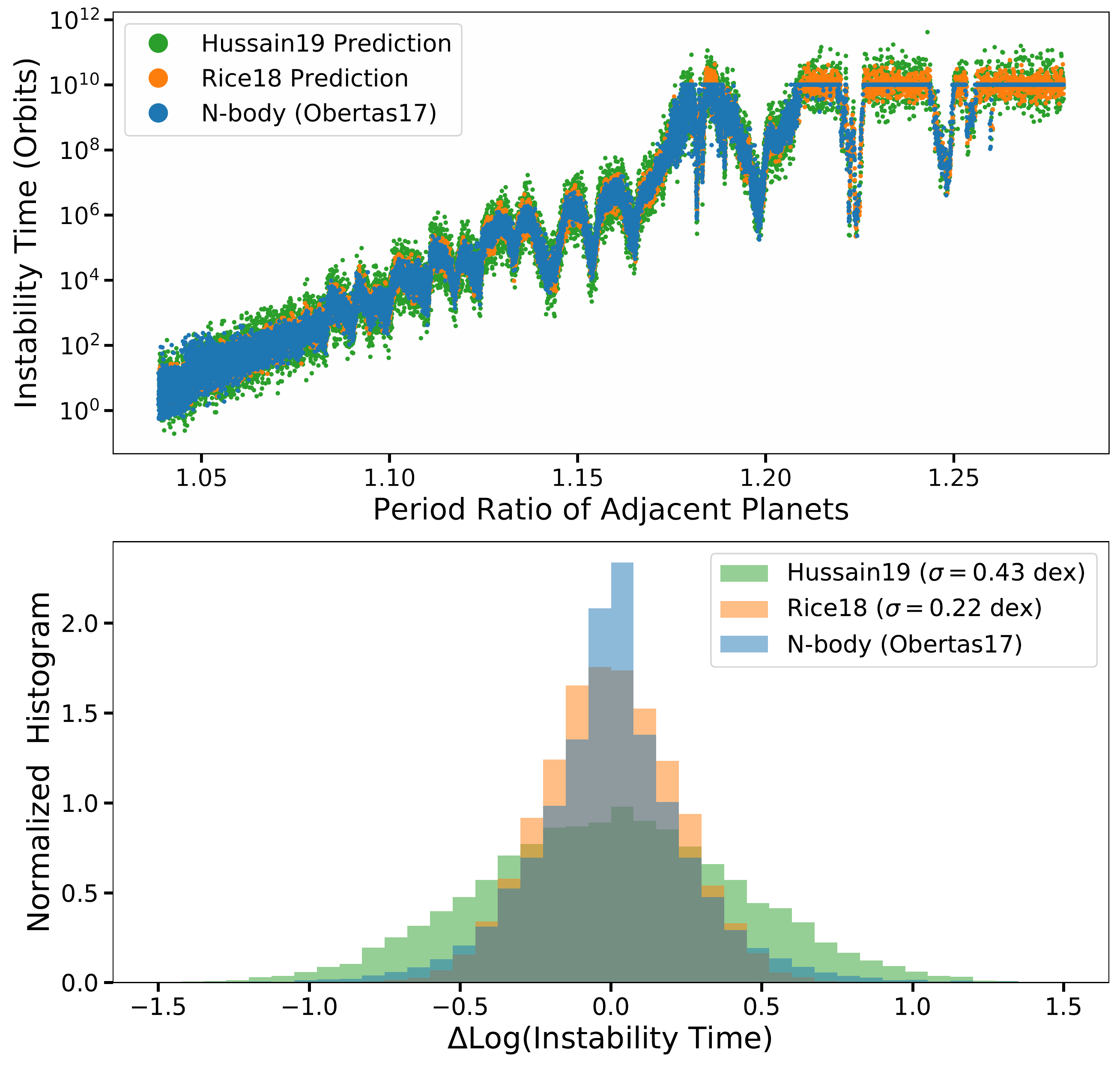}
    \caption{Top panel: Blue points denote instability times of co-planar, equal-mass, equal-separation planets on initially circular orbits from \protect\cite{Obertas17}, as a function of the period ratio of adjacent planets. Green points are drawn from a lognormal with standard deviation of $0.43$ dex (Sec.\:\ref{widthsec}) around the rolling median of instability times in a 10-sample window. Orange points are analogously drawn with the standard deviation of $0.22$ dex reported by \protect\cite{Rice18}.
    Standard deviations from the N-body integrations are approximately constant despite the instability times varying by ten orders of magnitude.
    Bottom panel: Histogram of the difference of log instability times plotted in the top panel away from the rolling median. The co-planar results of \protect\cite{Rice18} in orange match well with the co-planar N-body integrations in blue, which are more narrowly distributed than our 3-D results in green (see text).}
    \label{fig:obertas}
\end{figure}

While they ran no shadow integrations, the random choice in phases and fine sampling in separation achieves a similar effect.
On long timescales the phases can be averaged out and are unimportant, so in this sense, sampling randomly in phase would act as drawing an equivalent shadow trajectory\footnote{The phases in general do matter close to MMRs. The spreads in N-body instability times close to strong MMRs (i.e., in the sharp dips in top panel of Fig.\:\ref{fig:obertas}) are a factor of $\sim 2$ larger than in intervening regions, which we attribute to this effect; there we are not seeing the intrinsic width of instability times due to chaos, but rather that width convolved with a spread in initial conditions.\label{MMRfootnote}}.

The results from \cite{Obertas17} of log instability time vs the period ratio between adjacent planets are plotted in the top paneel of Fig.\:\ref{fig:obertas} as blue points.
More separated systems take longer to go unstable, and the sharp drops in instability times are at the location of MMRs \citep{Obertas17}.
Here we are interested in the spread of the trend.
In green we sample from our predicted lognormal with standard deviation of $0.43$ dex (Sec.\:\ref{widthsec}) around the rolling median of instability times in a 10-sample window.
In orange we sample instead from a lognormal with standard deviation $\approx 0.22$ dex, following \cite{Rice18}.

We see that our $0.43$ dex spread is typically too large, while the $0.22$ dex spread from \cite{Rice18} matches well.
This is easier to see in the bottom panel, where we plot a histogram of the difference of the log instability times plotted in the top panel away from the rolling median\footnote{We only include data up to a period ratio of 1.21 from the top panel, in order to avoid the long horizontal stretches, which are an artifact of the N-body integrations being cut off at $10^{10}$ orbits.}.
As discussed in Sec.\:\ref{widthsec}, we suspect that diffusion in the lower-dimensional planar configurations of \cite{Obertas17} and \cite{Rice18} yields narrower ITDs than our 3-D integrations.
The lognormal prediction of \cite{Rice18} would be a poor fit to the histogram in Fig.\:\ref{fig:comparing_shadow_systems} (twice as narrow as the green curve), and we compare the two results on two other 3-D configurations in the next two subsections.

We also note that while the match between the N-body integrations and the lognormal prediction from \cite{Rice18} is not exact, this is not a pressing concern, given that the N-body experiment of \cite{Obertas17} is not a perfect analogy to running shadow integrations (particularly near MMRs), and that the lognormal standard deviation of $0.22$ dex from \cite{Rice18} is derived from only three initial conditions.

Finally, leaving aside the {\it value} of the ITD standard deviation, the top panel of Fig.\:\ref{fig:obertas} shows that the ITD standard deviation remains constant, despite instability times varying by 10 orders of magnitude, even wider than the sample we analyzed above.
It is worth noting that this feature remains the same despite this test having almost double the number of planets, having equal mass planets and initial separations, and starting all planets on circular orbits.
The behavior may be changing in the right-most area of the plot, though it is difficult to say, given that the standard deviations are hard to measure where the instability times are changing so drastically, and the regions of interest are near strong MMRs \cite{Obertas17}, where our interpretation of nearby points as shadow trajectories breaks down (see footnote \ref{MMRfootnote}).
It may be interesting to further explore this region in future work.

\subsection{A Seven-Planet Resonant Chain}

As a test in the limit of high multiplicity that is strongly influenced by MMRs, we consider the case of TRAPPIST-1 \citep{Gillon17}, with seven $\sim$ Earth-mass planets in the longest resonant chain known to date, orbiting a late M-dwarf.
We consider a realization from the control sample in \cite{Tamayo17}, which generated randomly sampled initial conditions within the error bars reported by \cite{Gillon17}, and generated 500 shadows around this configuration by minutely shifting the third planet in the same way done above.
While the actual orbital inclinations in the system might be smaller \citep{Grimm18}, the inclinations relative to the sky plane in this test realization varied from $0.2-0.5^\circ$ across the planets.
Assuming randomly oriented orbital planes (node longitudes), this corresponds to mutual inclinations between adjacent planets of up to $0.8^\circ$, or vertical excursions of up to $80\%$ of a given planet's Hill radius.
This would suggest that the 3-D geometry is important, but it is not clearly in the non-planar regime.

\begin{figure}
    \centering
    \includegraphics[width = 0.5\textwidth]{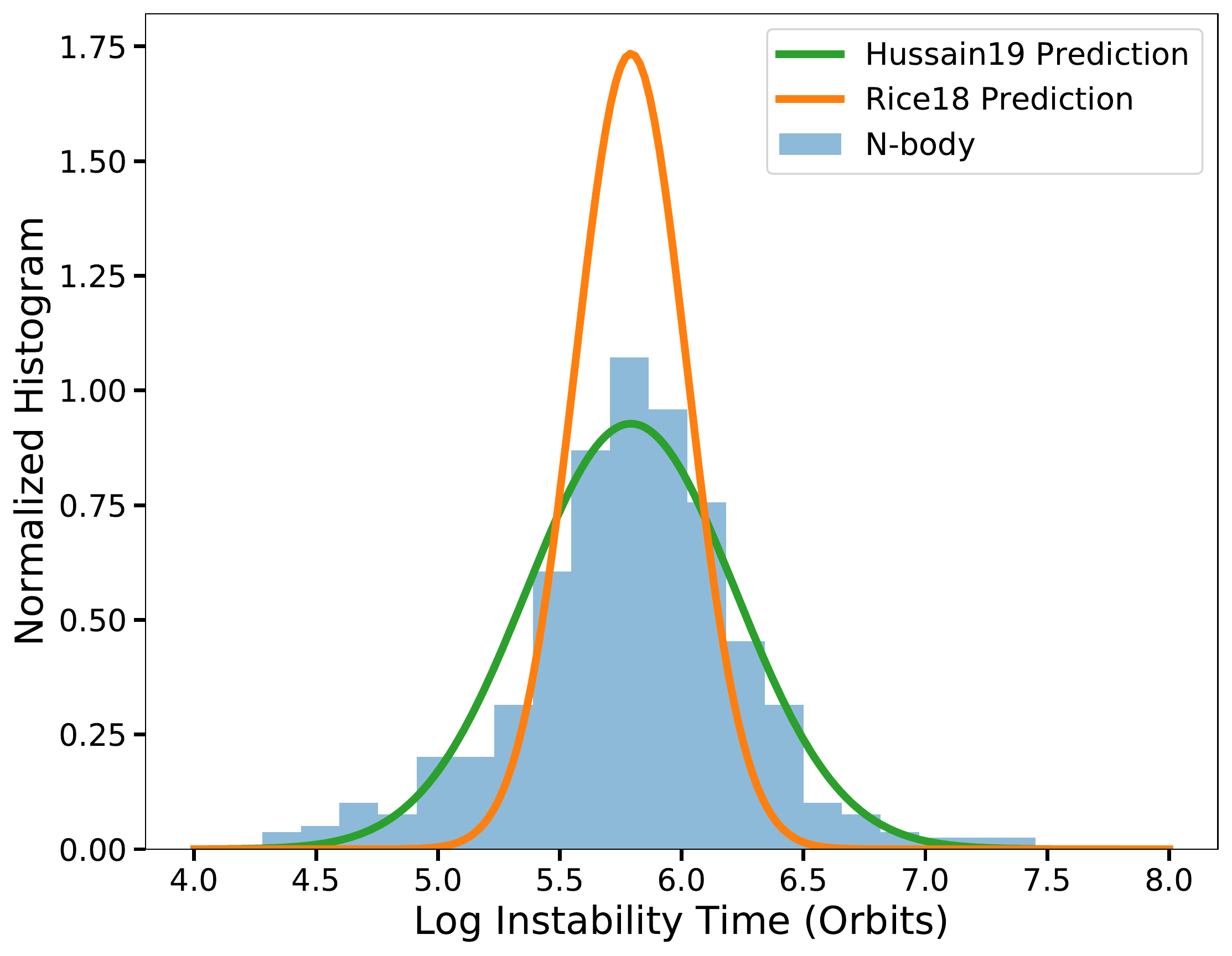}
    \caption{Blue histogram shows the distribution of instability times for a particular unstable configuration of the 7-planet resonant chain around TRAPPIST-1 (from \protect\cite{Tamayo17}). Green curve is our predicted lognormal distribution, which is statistically consistent with the N-body results (see text). Orange curve is the prediction reported from planar planetary systems by \protect\cite{Rice18}.  We argue in the text this is due to our adopting effectively 3-D configurations, while \protect\cite{Rice18} focused on effectively planar systems.}
    \label{fig:TRAPPIST_test}
\end{figure}

Figure \ref{fig:TRAPPIST_test} shows the resulting distribution of instability times, overlaid with our predicted lognormal with a standard deviation of $0.43$ (green), as well as the narrower prediction from \cite{Rice18}.
Both lognormals have been centered at the mean instability time of the blue histogram of the N-body results.
Computing a KS test between the TRAPPIST-1 ITD and an equal number of samples drawn from our prediction in green yields a p-value of 0.32, consistent with having been drawn from the same distribution.
By contrast, a KS comparison with the prediction from \cite{Rice18} yields a p-value of $4\times10^{-6}$.
 This further supports that our results extend to compact, high-multiplicity planetary systems.
It also reinforces that the deviation between our results and those of \cite{Rice18} likely corresponds to the difference between 2-D and 3-D planetary systems.
If the actual mutual inclinations in the system are much smaller, we would expect the ITD to follow \cite{Rice18}.

\subsection{Secular Systems}

Finally, we try pushing the boundaries of applicability for our predicted ITDs, and consider widely spaced systems like our Solar System.
Chaos and instability in our solar system is due to overlap of secular resonances \citep{Laskar90, Lithwick11secularchaos, Batygin15}, rather than overlap of MMRs, as likely governs tightly packed systems.
This is a very different regime to the one probed in all the above discussion, but one might wonder whether resonance overlap generically yields instability times with similar standard deviations.

The solar system is unstable on long timescales, with Mercury having a $\approx 1\%$ chance of colliding with the Sun or Venus in the rest of our star's main sequence lifetime \citep{Laskar09}.
For computational reasons, we artificially boost the eccentricities of all the planetary orbits, in order to increase the widths and overlaps of secular resonances, and hasten the onset of instability.
We first boost all orbital eccentricities by a factor of 1.45, and similarly stop the integrations when the Hill spheres of any pair of planets overlaps.
We run 200 shadow integrations.

\begin{figure}
    \centering
    \includegraphics[width= 0.5\textwidth]{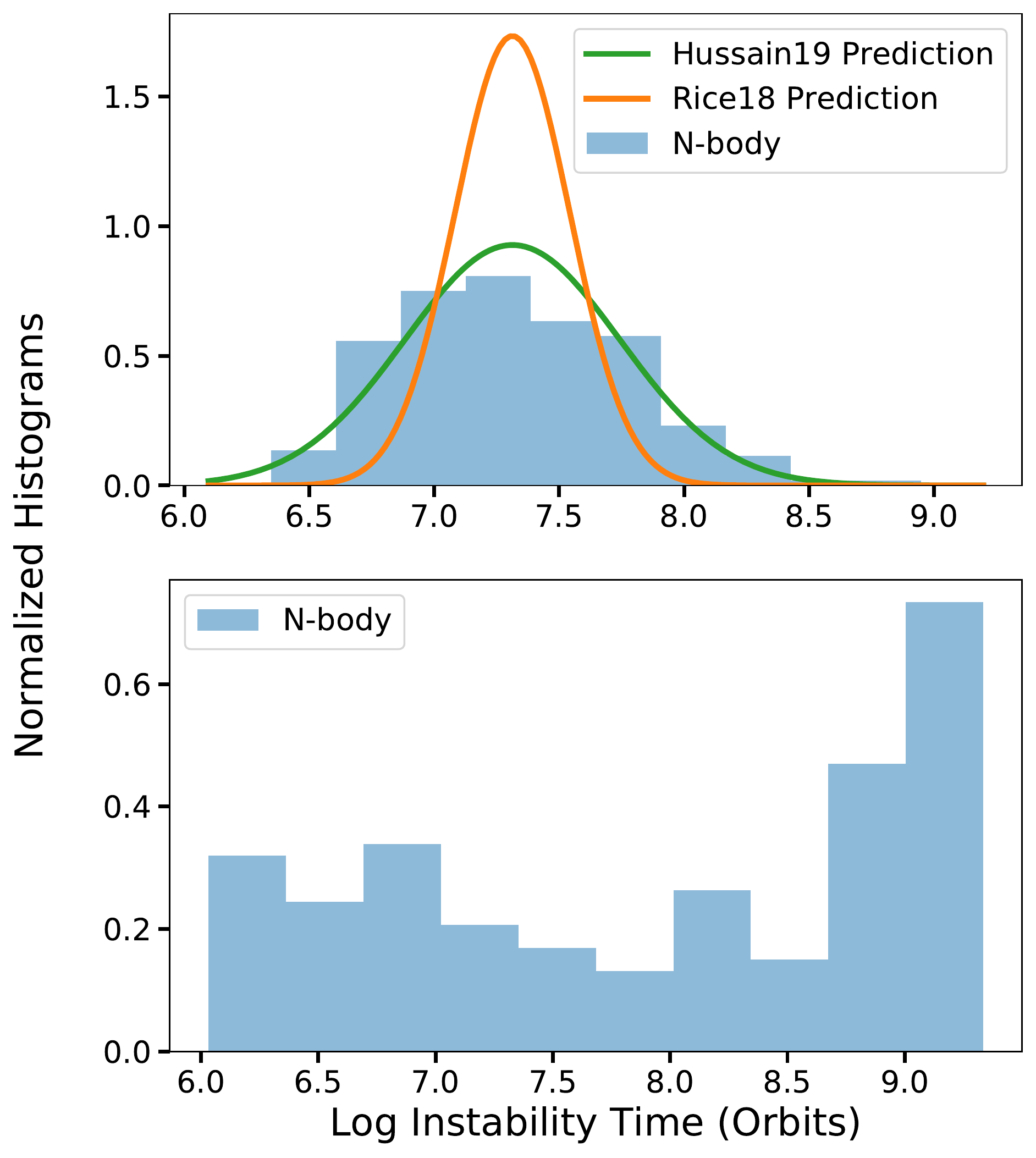}
    \caption{Distribution of solar systems instability times when all orbital eccentricities are initially increased by a factor of 1.45.
    Green curve is the same prediction as in Fig.\:\ref{fig:comparing_shadow_systems}.}
    \label{fig:solar_dist_1_45}.
\end{figure}

The resulting ITD is shown in the top panel of Fig.\:\ref{fig:solar_dist_1_45}, again overlaid with the same green lognormal from Fig.\:\ref{fig:TRAPPIST_test} with a standard deviation of $\approx 0.43$ dex, centered at the mean instability time of the ITD, as well as the narrower prediction from \cite{Rice18} in orange. Computing a KS test between the solar system ITD and an equal number of samples drawn from the green Gaussian distribution yields a p-value of $\approx$ 0.47, consistent with having been drawn from the same distribution.
Again, the results of \cite{Rice18} are too narrow, as one might expect given that the solar system is also 3-dimensional in the sense discussed above.
This would seem to suggest that our predicted ITDs are a generic outcome of chaotic transport due to resonance overlap, regardless of the types of resonances involved.

\label{discussion}

However, at lower eccentricities, the similarity is lost.
As seen in the bottom panel of Fig.\:\ref{fig:solar_dist_1_45}, if we boost the planetary eccentricities instead by a factor of 1.4, the resulting ITD is not at all lognormal (about $20\%$ of the integrations remained stable out to our cutoff time at $10^{9.5}$ Mercury orbits).

We interpret this as a difference between weak and strong chaos.
In the regime where resonances are strongly overlapped, and chaotic regions fill large fractions of phase space, diffusive random walks are good models for the chaotic transport in action space \citep[e.g.][]{Lichtenberg13}.
This is likely the typical regime for the very tightly packed systems analyzed above, and when solar system orbital eccentricities are boosted by large factors.
For more widely separated systems at lower eccentricities, chaos is limited to thin layers around the resonance separatrices, and most of the phase space is stable over very long timescales.
At the transition toward instability, we speculate that even though systems can explore the chaotic web and eventually reach instability, diffusive random walks are poor models for the chaotic transport, with strong boundary effects, e.g., systems getting stuck for long periods near resonance separatrices.
We suspect that the approach to a common, well defined lognormal distribution of instability times is the generic result of exploring strongly chaotic regions where nearby trajectories get vigorously mixed.
We plan to explore and test this idea in future work.

While the above results suggest caution, it is nevertheless interesting to apply our predictions to the real solar system.
Given that \cite{Laskar09} found that Mercury went unstable in $\approx 1\%$ of realizations within 5 Gyr (about a $2-\sigma$ result), our analysis would suggest that the mean instability time for Mercury should be a factor of $10^{2 \times 0.43}$ longer, $\approx 40$ Gyr.
While this is far beyond the Sun's main sequence lifetime (at which point Mercury would become engulfed), this hypothetical mean instability time is an interesting constraint on how close to the edge of instability the planet formation process left our solar system.
It would be valuable (though computationally costly!) to test this with direct N-body integrations.

\section{Conclusion} \label{conclusion}

Instability times measured through direct N-body integration are not exact; rather, they should be viewed as draws from a distribution of possible outcomes.
When regular dynamics takes the system onto orbit-crossing configurations, instability time distributions (ITDs) will be sharply peaked, since nearby initial conditions won't have time to diverge from one another (e.g., bottom left panel of Fig.\:\ref{outliers}, top panels of Fig.\:\ref{hists}).
However, in the limit of long random walks where many Lyapunov (chaotic) timescales elapse before instability, nearby trajectories will have time to chaotically mix and instability times can settle into well-defined statistical distributions (e.g., bottom right panel of Fig.\:\ref{outliers}, bottom panels of Fig.\:\ref{hists}).

From Fig.\:\ref{outliers}, this is typically the case for systems with instability times beyond $\sim10^5$ orbits, though it also applies to many configurations with shorter instability times.
For instability times below $10^5$ orbits, one can calculate the Lyapunov time as done in Fig.\:\ref{outliers} to identify configurations whose shadow trajectories do not chaotically mix sufficiently to generate lognormal ITDs; alternatively, it is computationally trivial for such short integrations to rerun slightly offset initial conditions to test for this directly.
We find that the limiting ITDs of long random walks are lognormal, and that the standard deviations of these lognormal distributions are approximately constant, despite typical instability times varying from $10^4-10^8$ orbits across the planetary configurations in our sample.
We argue that the standard deviation of these lognormals depends on whether or not the system is effectively co-planar, i.e., whether bodies vertical excursions are much smaller than their Hill spheres.
\cite{Rice18} report a standard deviation of $0.22$ dex from a sample of three effectively planar initial conditions, which matches well with the 20,000 co-planar integrations of \cite{Obertas17} (Fig.\:\ref{fig:obertas}).
For a sample of 454 non-coplanar configurations with instability times varying from $10^4-10^8$, we find lognormal standard deviations sharply peaked around $0.43$ dex $ \pm 0.16$  (Fig.\:\ref{fig:comparing_sigmas}).
This allowed us to quantitatively evaluate two possible definitions on the uncertainty in instability times measured from N-body integrations:
\begin{itemize}
    \item Given a single instability time measurement $t$, the mean of the instability time distribution will be lognormally distributed around $t$ with a standard deviation of $\approx 0.4$ dex (0.2 dex for planar configurations). This corresponds to a Gaussian error of 0.4 (0.2) in the exponent of the instability time, or a multiplicative factor of approximately 2.7 (1.7) in either direction.
    \item Given a single instability time measurement $t$, the distribution for additional measurements $t'$ will be lognormally distributed around $t$ with a standard deviation wider by a factor of $\sqrt{2}$, $\approx 0.6$ dex (0.3 dex for planar configurations). This corresponds to a factor of 4 (2.1) in either direction for the absolute instability time, for 3-D (2-D) configurations, respectively. This reflects the intrinsic uncertainty generated by chaos (even in the case where the mean of the ITD is known), combined with the fact that we do not know where the mean of the ITD is (as quantified in the bullet above).
\end{itemize}
These uncertainties define the fundamental limit imposed by chaos on the predictability of instability times in tightly packed planetary systems.
Typically the first definition above would be the appropriate intrinsic uncertainty to quote on a measured instability time, roughly a factor of 3 in either direction.

We find excellent agreement testing these predictions derived from $454$ closely packed configurations on a much wider validation set of $\approx 10,000$ integrations in Fig.\:\ref{fig:comparing_shadow_systems} and on a seven-planet resonant chain around TRAPPIST-1 in Fig.\:\ref{fig:TRAPPIST_test}.
Finally, we pushed beyond tightly packed systems to realizations of the solar system with artificially increased orbital eccentricities, finding good agreement in strongly chaotic cases, but finding significant deviations near the boundaries of instability.
We speculate that this behavior is related to complicated correlations when traversing thin chaotic layers in phase space near the stability boundary.

The statistical characterization of ITDs presented in this paper provide important constraints on dynamical models of chaotic transport in compact planetary systems. It would be valuable to derive the parameters of such random walks from first principles, and to further explore the range of applicability of the results presented here.
Toward this end we have made all \reb {\:\tt SimulationArchives} \citep{ReinTamayo17} analyzed in this paper publicly available at \url{https://zenodo.org/record/3461173}.
We provide instructions and the scripts necessary to reproduce all the figures above at \url{https://github.com/Naireen/StabilitySetImage}.

\section*{Acknowledgments}
We are grateful to the anonymous reviewer for an insightful review that greatly strengthened this manuscript.
We would also like to thank Hanno Rein and Alysa Obertas for insightful discussions during the preparation of this manuscript. We would also like to thank Alysa Obertas for sharing her integration results plotted in Fig. 5, and Sam Hadden for all his help developing the {\tt celmech} package that we used to initialize resonant systems. Support for this work was provided by NASA through the NASA Hubble Fellowship grant HST-HF2-51423.001-A awarded  by  the  Space  Telescope  Science  Institute,  which  is  operated  by  the  Association  of  Universities  for  Research  in  Astronomy,  Inc.,  for  NASA,  under  contract  NAS5-26555.
This research was made possible by the open-source projects \texttt{Jupyter} \citep{jupyter}, \texttt{iPython} \citep{ipython}, and \texttt{matplotlib} \citep{matplotlib, matplotlib2}.

\bibliography{ms}

\begin{thebibliography}{}
\makeatletter
\relax
\def\mn@urlcharsother{\let\do\@makeother \do\$\do\&\do\#\do\^\do\_\do\%\do\~}
\def\mn@doi{\begingroup\mn@urlcharsother \@ifnextchar [ {\mn@doi@}
  {\mn@doi@[]}}
\def\mn@doi@[#1]#2{\def\@tempa{#1}\ifx\@tempa\@empty \href
  {http://dx.doi.org/#2} {doi:#2}\else \href {http://dx.doi.org/#2} {#1}\fi
  \endgroup}
\def\mn@eprint#1#2{\mn@eprint@#1:#2::\@nil}
\def\mn@eprint@arXiv#1{\href {http://arxiv.org/abs/#1} {{\tt arXiv:#1}}}
\def\mn@eprint@dblp#1{\href {http://dblp.uni-trier.de/rec/bibtex/#1.xml}
  {dblp:#1}}
\def\mn@eprint@#1:#2:#3:#4\@nil{\def\@tempa {#1}\def\@tempb {#2}\def\@tempc
  {#3}\ifx \@tempc \@empty \let \@tempc \@tempb \let \@tempb \@tempa \fi \ifx
  \@tempb \@empty \def\@tempb {arXiv}\fi \@ifundefined
  {mn@eprint@\@tempb}{\@tempb:\@tempc}{\expandafter \expandafter \csname
  mn@eprint@\@tempb\endcsname \expandafter{\@tempc}}}

\bibitem[\protect\citeauthoryear{{Batygin}, {Morbidelli}  \&
  {Holman}}{{Batygin} et~al.}{2015}]{Batygin15}
{Batygin} K.,  {Morbidelli} A.,   {Holman} M.~J.,  2015, \mn@doi [\apj]
  {10.1088/0004-637X/799/2/120}, \href
  {http://adsabs.harvard.edu/abs/2015ApJ...799..120B} {799, 120}

\bibitem[\protect\citeauthoryear{Carter et~al.,}{Carter
  et~al.}{2012}]{Carter12}
Carter J.~A.,  et~al., 2012, Science, 337, 556

\bibitem[\protect\citeauthoryear{{Chambers}, {Wetherill}  \& {Boss}}{{Chambers}
  et~al.}{1996}]{Chambers96}
{Chambers} J.~E.,  {Wetherill} G.~W.,   {Boss} A.~P.,  1996, \mn@doi [Icarus]
  {10.1006/icar.1996.0019}, \href
  {http://adsabs.harvard.edu/abs/1996Icar..119..261C} {119, 261}

\bibitem[\protect\citeauthoryear{{Chatterjee}, {Ford}, {Matsumura}  \&
  {Rasio}}{{Chatterjee} et~al.}{2008}]{Chatterjee08}
{Chatterjee} S.,  {Ford} E.~B.,  {Matsumura} S.,   {Rasio} F.~A.,  2008,
  \mn@doi [\apj] {10.1086/590227}, \href
  {http://adsabs.harvard.edu/abs/2008ApJ...686..580C} {686, 580}

\bibitem[\protect\citeauthoryear{Chirikov}{Chirikov}{1979}]{Chirikov79}
Chirikov B.~V.,  1979, Physics reports, 52, 263

\bibitem[\protect\citeauthoryear{Deck \& Agol}{Deck \& Agol}{2015}]{Deck15}
Deck K.~M.,  Agol E.,  2015, The Astrophysical Journal, 802, 116

\bibitem[\protect\citeauthoryear{Droettboom et~al.,}{Droettboom
  et~al.}{2016}]{matplotlib2}
Droettboom M.,  et~al., 2016, matplotlib: matplotlib v1.5.1,
  \mn@doi{10.5281/zenodo.44579}, \url {http://dx.doi.org/10.5281/zenodo.44579}

\bibitem[\protect\citeauthoryear{Faber \& Quillen}{Faber \&
  Quillen}{2007}]{Faber07}
Faber P.,  Quillen A.~C.,  2007, Monthly Notices of the Royal Astronomical
  Society, 382, 1823

\bibitem[\protect\citeauthoryear{Fabrycky et~al.,}{Fabrycky
  et~al.}{2014}]{Fabrycky14}
Fabrycky D.~C.,  et~al., 2014, The Astrophysical Journal, 790, 146

\bibitem[\protect\citeauthoryear{Foreman-Mackey, Hogg, Lang  \&
  Goodman}{Foreman-Mackey et~al.}{2013}]{Foreman13}
Foreman-Mackey D.,  Hogg D.~W.,  Lang D.,   Goodman J.,  2013, Publications of
  the Astronomical Society of the Pacific, 125, 306

\bibitem[\protect\citeauthoryear{Funk, Wuchterl, Schwarz, Pilat-Lohinger  \&
  Eggl}{Funk et~al.}{2010}]{Funk10}
Funk B.,  Wuchterl G.,  Schwarz R.,  Pilat-Lohinger E.,   Eggl S.,  2010,
  Astronomy \& Astrophysics, 516, A82

\bibitem[\protect\citeauthoryear{Gillon et~al.,}{Gillon
  et~al.}{2017}]{Gillon17}
Gillon M.,  et~al., 2017, Nature, 542, 456

\bibitem[\protect\citeauthoryear{{Gladman}}{{Gladman}}{1993}]{Gladman93}
{Gladman} B.,  1993, \mn@doi [Icarus] {10.1006/icar.1993.1169}, \href
  {http://adsabs.harvard.edu/abs/1993Icar..106..247G} {106, 247}

\bibitem[\protect\citeauthoryear{Grimm et~al.,}{Grimm et~al.}{2018}]{Grimm18}
Grimm S.~L.,  et~al., 2018, Astronomy \& Astrophysics, 613, A68

\bibitem[\protect\citeauthoryear{Hadden}{Hadden}{2019}]{Hadden19}
Hadden S.,  2019, arXiv preprint arXiv:1909.05264

\bibitem[\protect\citeauthoryear{Hadden \& Lithwick}{Hadden \&
  Lithwick}{2014}]{Hadden14}
Hadden S.,  Lithwick Y.,  2014, The Astrophysical Journal, 787, 80

\bibitem[\protect\citeauthoryear{Hadden \& Lithwick}{Hadden \&
  Lithwick}{2016}]{Hadden16}
Hadden S.,  Lithwick Y.,  2016, The Astrophysical Journal, 828, 44

\bibitem[\protect\citeauthoryear{Hadden \& Lithwick}{Hadden \&
  Lithwick}{2017}]{Hadden17}
Hadden S.,  Lithwick Y.,  2017, The Astronomical Journal, 154, 5

\bibitem[\protect\citeauthoryear{Hadden \& Lithwick}{Hadden \&
  Lithwick}{2018}]{Hadden18}
Hadden S.,  Lithwick Y.,  2018, The Astronomical Journal, 156, 95

\bibitem[\protect\citeauthoryear{Hunter}{Hunter}{2007}]{matplotlib}
Hunter J.~D.,  2007, Computing In Science \& Engineering, 9, 90

\bibitem[\protect\citeauthoryear{Izidoro, Ogihara, Raymond, Morbidelli,
  Pierens, Bitsch, Cossou  \& Hersant}{Izidoro et~al.}{2017}]{Izidoro17}
Izidoro A.,  Ogihara M.,  Raymond S.~N.,  Morbidelli A.,  Pierens A.,  Bitsch
  B.,  Cossou C.,   Hersant F.,  2017, Monthly Notices of the Royal
  Astronomical Society, 470, 1750

\bibitem[\protect\citeauthoryear{Jontof-Hutter, Lissauer, Rowe  \&
  Fabrycky}{Jontof-Hutter et~al.}{2014}]{Jontof14}
Jontof-Hutter D.,  Lissauer J.~J.,  Rowe J.~F.,   Fabrycky D.~C.,  2014, The
  Astrophysical Journal, 785, 15

\bibitem[\protect\citeauthoryear{Kluyver et~al.,}{Kluyver
  et~al.}{2016}]{jupyter}
Kluyver T.,  et~al., 2016, Positioning and Power in Academic Publishing:
  Players, Agents and Agendas, p.~87

\bibitem[\protect\citeauthoryear{Lam \& Kipping}{Lam \& Kipping}{2018}]{Lam18}
Lam C.,  Kipping D.,  2018, Monthly Notices of the Royal Astronomical Society,
  476, 5692

\bibitem[\protect\citeauthoryear{Laskar}{Laskar}{1990}]{Laskar90}
Laskar J.,  1990, Icarus, 88, 266

\bibitem[\protect\citeauthoryear{Laskar \& Gastineau}{Laskar \&
  Gastineau}{2009}]{Laskar09}
Laskar J.,  Gastineau M.,  2009, Nature, 459, 817

\bibitem[\protect\citeauthoryear{Lichtenberg \& Lieberman}{Lichtenberg \&
  Lieberman}{2013}]{Lichtenberg13}
Lichtenberg A.~J.,  Lieberman M.~A.,  2013, Regular and stochastic motion.
 Vol. 38, Springer Science \& Business Media

\bibitem[\protect\citeauthoryear{Lissauer et~al.,}{Lissauer
  et~al.}{2011}]{Lissauer11}
Lissauer J.~J.,  et~al., 2011, The Astrophysical Journal Supplement Series,
  197, 8

\bibitem[\protect\citeauthoryear{{Lithwick} \& {Wu}}{{Lithwick} \&
  {Wu}}{2011}]{Lithwick11secularchaos}
{Lithwick} Y.,  {Wu} Y.,  2011, \mn@doi [\apj] {10.1088/0004-637X/739/1/31},
  \href {http://adsabs.harvard.edu/abs/2011ApJ...739...31L} {739, 31}

\bibitem[\protect\citeauthoryear{{Marchal} \& {Bozis}}{{Marchal} \&
  {Bozis}}{1982}]{Marchal82}
{Marchal} C.,  {Bozis} G.,  1982, \mn@doi [Celestial Mechanics]
  {10.1007/BF01230725}, \href
  {http://adsabs.harvard.edu/abs/1982CeMec..26..311M} {26, 311}

\bibitem[\protect\citeauthoryear{{Marzari} \& {Weidenschilling}}{{Marzari} \&
  {Weidenschilling}}{2002}]{Marzari02}
{Marzari} F.,  {Weidenschilling} S.~J.,  2002, \mn@doi [Icarus]
  {10.1006/icar.2001.6786}, \href
  {http://adsabs.harvard.edu/abs/2002Icar..156..570M} {156, 570}

\bibitem[\protect\citeauthoryear{Murray \& Holman}{Murray \&
  Holman}{1997}]{Murray97}
Murray N.,  Holman M.,  1997, The Astronomical Journal, 114, 1246

\bibitem[\protect\citeauthoryear{Obertas, Van~Laerhoven  \& Tamayo}{Obertas
  et~al.}{2017}]{Obertas17}
Obertas A.,  Van~Laerhoven C.,   Tamayo D.,  2017, \mn@doi [Icarus]
  {10.1016/j.icarus.2017.04.010}, 293, 52

\bibitem[\protect\citeauthoryear{Pepe et~al.,}{Pepe et~al.}{2013}]{Pepe13}
Pepe F.,  et~al., 2013, Nature, 503, 377

\bibitem[\protect\citeauthoryear{P\'erez \& Granger}{P\'erez \&
  Granger}{2007}]{ipython}
P\'erez F.,  Granger B.~E.,  2007, \mn@doi [Computing in Science and
  Engineering] {10.1109/MCSE.2007.53}, 9, 21

\bibitem[\protect\citeauthoryear{Pu \& Wu}{Pu \& Wu}{2015}]{Pu15}
Pu B.,  Wu Y.,  2015, The Astrophysical Journal, 807, 44

\bibitem[\protect\citeauthoryear{Quarles, Quintana, Lopez, Schlieder  \&
  Barclay}{Quarles et~al.}{2017}]{Quarles17}
Quarles B.,  Quintana E.~V.,  Lopez E.~D.,  Schlieder J.~E.,   Barclay T.,
  2017, arXiv preprint arXiv:1704.02261

\bibitem[\protect\citeauthoryear{Quillen}{Quillen}{2011}]{Quillen11}
Quillen A.~C.,  2011, Monthly Notices of the Royal Astronomical Society, 418,
  1043

\bibitem[\protect\citeauthoryear{Rein \& Liu}{Rein \& Liu}{2012}]{Rein12}
Rein H.,  Liu S.-F.,  2012, Astronomy \& Astrophysics, 537, A128

\bibitem[\protect\citeauthoryear{Rein \& Tamayo}{Rein \&
  Tamayo}{2015}]{ReinTamayo15}
Rein H.,  Tamayo D.,  2015, Monthly Notices of the Royal Astronomical Society,
  452, 376

\bibitem[\protect\citeauthoryear{Rein \& Tamayo}{Rein \&
  Tamayo}{2017}]{ReinTamayo17}
Rein H.,  Tamayo D.,  2017, \mn@doi [Monthly Notices of the Royal Astronomical
  Society] {10.1093/mnras/stx232}, 467, 2377

\bibitem[\protect\citeauthoryear{Rice, Rasio  \& Steffen}{Rice
  et~al.}{2018}]{Rice18}
Rice D.~R.,  Rasio F.~A.,   Steffen J.~H.,  2018, Monthly Notices of the Royal
  Astronomical Society, 481, 2205

\bibitem[\protect\citeauthoryear{Smith \& Lissauer}{Smith \&
  Lissauer}{2009}]{Smith09}
Smith A.~W.,  Lissauer J.~J.,  2009, Icarus, 201, 381

\bibitem[\protect\citeauthoryear{Steffen et~al.,}{Steffen
  et~al.}{2013}]{Steffen13}
Steffen J.~H.,  et~al., 2013, Monthly Notices of the Royal Astronomical
  Society, 428, 1077

\bibitem[\protect\citeauthoryear{{Tamayo}, {Hedman}  \& {Burns}}{{Tamayo}
  et~al.}{2014}]{Tamayo14a}
{Tamayo} D.,  {Hedman} M.~M.,   {Burns} J.~A.,  2014, \mn@doi [Icarus]
  {10.1016/j.icarus.2014.01.021}, \href
  {http://adsabs.harvard.edu/abs/2014Icar..233....1T} {233, 1}

\bibitem[\protect\citeauthoryear{Tamayo, Triaud, Menou  \& Rein}{Tamayo
  et~al.}{2015}]{Tamayo15}
Tamayo D.,  Triaud A.~H.,  Menou K.,   Rein H.,  2015, The Astrophysical
  Journal, 805, 100

\bibitem[\protect\citeauthoryear{Tamayo et~al.,}{Tamayo
  et~al.}{2016}]{Tamayo16}
Tamayo D.,  et~al., 2016, The Astrophysical Journal Letters, 832, L22

\bibitem[\protect\citeauthoryear{Tamayo, Rein, Petrovich  \& Murray}{Tamayo
  et~al.}{2017}]{Tamayo17}
Tamayo D.,  Rein H.,  Petrovich C.,   Murray N.,  2017, The Astrophysical
  Journal Letters, 840, L19

\bibitem[\protect\citeauthoryear{Volk \& Gladman}{Volk \&
  Gladman}{2015}]{Volk15}
Volk K.,  Gladman B.,  2015, The Astrophysical Journal Letters, 806, L26

\bibitem[\protect\citeauthoryear{Wang et~al.,}{Wang et~al.}{2018}]{Wang18}
Wang J.~J.,  et~al., 2018, The Astronomical Journal, 156, 192

\bibitem[\protect\citeauthoryear{Wisdom}{Wisdom}{1980}]{Wisdom80}
Wisdom J.,  1980, The Astronomical Journal, 85, 1122

\bibitem[\protect\citeauthoryear{Wu, Zhang, Zhou  \& Steffen}{Wu
  et~al.}{2019}]{Wu19}
Wu D.-H.,  Zhang R.~C.,  Zhou J.-L.,   Steffen J.~H.,  2019, Monthly Notices of
  the Royal Astronomical Society, 484, 1538

\bibitem[\protect\citeauthoryear{Yoshinaga, Kokubo  \& Makino}{Yoshinaga
  et~al.}{1999}]{Yoshinaga99}
Yoshinaga K.,  Kokubo E.,   Makino J.,  1999, Icarus, 139, 328

\bibitem[\protect\citeauthoryear{Zhou, Lin  \& Sun}{Zhou et~al.}{2007}]{Zhou07}
Zhou J.-L.,  Lin D.~N.,   Sun Y.-S.,  2007, The Astrophysical Journal, 666, 423

\makeatother
\end{thebibliography}

\appendix

\section{Dataset} \label{dataset}

The dataset of Tamayo et al. ({\it in prep}) will soon be made publicly available online, and a detailed analysis published.
In order for this work to stand alone, we provide an overview of the setup.
As mentioned in the main text, there is both a `random' and `resonant' dataset, but before discussing their differences we mention their commonalities.

In both datasets, orbits had inclinations log-uniformly and independently sampled in the range from $10^{-3}-10^{-1}$ radians, and the longitudes of node uniformly drawn from $[0,2\pi]$.
This corresponds to maximum mutual inclinations of $\approx 11^\circ$.
These inclinations are comparable to the integrations of \cite{Chatterjee08} (uniformly sampled between $[0,10^\circ]$), but much larger than those of \cite{Rice18}, which are effectively planar (with Hill spheres much larger than the vertical orbital excursions given by the product of the semimajor axis and inclination).
Most previous studies draw from Rayleigh distributions.
We do not expect this to have a strong effect on stability, but the choice to sample log-normally by Tamayo et al. was made to cover parameter space agnostically, and not restrict the analysis to a particular scale.

The mass ratios of all three planets relative to the central star were chosen independently from a log-uniform distribution, between $10^{-7}$ ($\sim1/3$ Mars) and $10^{-4}$ ($\sim2$x Neptune).
This is in the range probed by \cite{Rice18}, who used fixed mass ratios of $10^{-5}$, but smaller than the range studied by \cite{Chatterjee08}, in the giant planet range of a few times $10^{-4}$ to a few times $10^{-3}$.
These differences in masses and inclinations have consequences that we discuss in the main text.
We also note that while we run three-planet systems like \cite{Chatterjee08}, \cite{Rice18} run four-planet systems; while there is a qualitative difference between the two and three-planet case, results at multiciplicities of three and higher are qualitatively similar \citep{Chambers96}.

All integrations were performed with \whfast \citep{ReinTamayo15}, part of the \reb N-body package \citep{Rein12}.
All cases use a timestep of $\approx 3.4\%$ the innermost planet's orbital period.
The simulation was stopped and the instability time recorded if any planets' Hill spheres overlap.
The specific halting condition is not important \citep{Gladman93}; once Hill spheres start crossing, the system becomes an orbit-crossing tangle on orbital timescales\footnote{It might still take a long time for small planets close to their host star to find one another and collide \citep{Rice18}; however, in the context of applying stability constraints, we are usually interested in the time to instability defined such that the system architecture becomes inconsistent with the typically observed multi-planet system with approximately planar, concentric near-circular orbits.}.

The integrations analyzed in this work were saved in the \reb {\:\tt SimulationArchive} format, which enables exact, machine-independent reproducibility of results \citep{ReinTamayo17}.
This is particularly valuable for strongly chaotic systems like the ones analyzed in this work.
All {\tt SimulationArchives} are publicly available at \url{https://zenodo.org/record/3461173}, and we provide instructions and the scripts necessary to reproduce the figures in this paper at \url{https://github.com/Naireen/StabilitySetImage}.

\subsection{Random Dataset} \label{randdataset}
The `random' set of initial conditions was generated by uniformly sampling the semimajor axis separation between adjacent planets independently, in the range between [0,30] mutual Hill radii.
For the mass ratios we consider, this corresponds to keeping the period ratios between adjacent planets $\lesssim 1.5$, comparable to \cite{Rice18}, but narrower than for the giant planets in \cite{Chatterjee08}.
We comment on these implications below.
Eccentricities were drawn log-uniformly between the characteristic eccentricities imparted at conjunctions (taken as the ratio of the interplanetary forces to the central force from the star) and the orbit-crossing value.
Pericenter orientations and phases were drawn uniformly from $[0, 2\pi]$.

We drew at random $\approx 280$ systems with instability times between $10^4-10^7$ orbits from the wider dataset of Tamayo et al. ({\it in prep.}). We then ran 1000 shadow trajectories for $10^8$ orbits for each of these initial conditions. Finally, we selected systems with mean instability times between $10^4-10^7$ orbits, leaving 246 systems in our random sample.

\subsection{Resonant Dataset} \label{resdataset}
The `resonant' set of initial conditions was generated by randomly choosing a pair of planets (inner pair, outer pair, or non-adjacent), putting them in or near resonance, and then choosing the third planet randomly as above, within 30 Hill radii. In other words, two of the planets are strongly influenced by a strong MMR, but are not necessarily resonantly interacting with the third planet. We initialized the resonant pair with the open-source {\tt celmech} package (\url{https://github.com/shadden/celmech}), which includes an API for generating orbital configurations \reb {\:\tt Simulations} from a set of resonant parameters, based on \cite{Hadden19}.

In particular, for the resonant pair of planets, the pair was initialized in one of the randomly chosen first (n:n-1) or second (n:n-2) order MMR in the range from [3.5, 30] mutual Hill radii, where the lower limit is the Hill stability limit \citep{Gladman93}, and was chosen to avoid immediate instabilities.
The equilibrium eccentricity forced by the resonance (equivalently, the depth in resonance) was sampled log-uniformly from the value induced by the planets on one another at conjunction to the orbit-crossing value, and the initial distance from the equilibrium eccentricity was sampled log-uniformly from $[3\times10^{-3}, 3]$ times the distance to the separatrix, the boundary of the resonance where the dynamics will be most chaotic. Thus, this resonantly initialized pair of planets spans the range from being in resonance to being outside the resonant region, but still having their dynamics strongly influenced by the MMR.
The latter is the region inhabited by many of the planets discovered by the Kepler mission to exhibit transit timing variations \citep[e.g.,][]{Hadden16}.

After initializing the resonant pair, the third planet is added following the procedure for the random systems.
We note that the presence of this third planet means that the two-body resonance parameters like the eccentricity forced by the resonance are no longer exact, but this procedure ensures that the MMR strongly influences the dynamics, and in most cases it is indeed the dominant perturbation.
For details, see the companion paper by Tamayo et al., {\it in prep}.

For this case we drew $\approx 200$ systems with instability times between $10^4-10^7$ orbits. For computational reasons, we only ran 500 shadow trajectories for each initial condition, also for $10^8$ orbits. We then selected systems with mean instability times between $10^4-10^7$ orbits, leaving 208 systems in the resonant sample. We note that while the random systems were drawn from the corresponding wider set of Tamayo et al. ({\it in prep.}), for idiosyncratic reasons the resonant system were generated independently, using the same script as the wider resonant sample in Tamayo et al. ({\it in prep.}), but with different random seeds.

\end{document}